\def\sp{\hspace{1.5pt}}
\def\etal{{\it et~al.}}
\def\amin{\ifmmode^{\prime}\else$^{\prime}$\fi}
\def\asec{\ifmmode^{\prime\prime}\else$^{\prime\prime}$\fi}

\def\simgt{\lower.5ex\hbox{$\; \buildrel > \over \sim \;$}}
\def\simlt{\lower.5ex\hbox{$\; \buildrel < \over \sim \;$}}

\newcommand\ASCA{{\it ASCA}}
\newcommand\asca{{\it ASCA}}

\def\sp{\hskip 1.5pt}
\def\psr{\hbox{AX{\sp}J1845$-$0258}}
\def\src{\hbox{AX{\sp}J184453.3$-$025642}}
\def\snr{G29.6+0.1}

\documentstyle[bologna,epsfig]{article}

\title{A New Pulsar/SNR Pair: \psr\ in G29.6+0.1}

\author{E.V.~Gotthelf$^1$, G.~Vasisht$^2$, K.~Torii$^3$ \& B.M.~Gaensler$^4$}

\affil{1) Columbia University, 550
West 120th Street, New York, NY 10027 \\ 2) 
California Institute of Technology, 4800 Oak Grove Drive, Pasadena, CA
91109 \\ 3) NASDA TKSC SURP, 2-1-1 Sengen, Tsukuba, Ibaraki 305-8505, Japan \\ 4) Massachusetts Institute of
Technology, 70 Vassar Street, Cambridge, MA 02139}

\begin{document}

\maketitle

\begin{abstract}

We present a follow-up X-ray and radio study of the field containing
the 7-s X-ray pulsar \psr, the serendipitous \asca\ source whose
characteristics are found to be similar to those of the anomalous
X-ray pulsars (AXPs). Newly acquired \ASCA\ data confirms a dramatic
reduction in flux from the pulsar and reveals instead a faint
X-ray point source, \src, within the pulsar's error circle. This X-ray
source is surrounded by a partial shell of emission coincident with a
newly discovered young shell-type radio supernova remnant, \snr. The
central X-ray source is too faint to provide a detection of the
expected pulsations which might confirm \src\ as the pulsar. We argue
that this system is similar to that of RCW~103, another AXP-like
object whose central source displays low/high flux states (but no
pulsations). The alternative interpretation of a binary system,
perhaps associated with a supernova remnant, is still possible. In
either case, this result may have profound implication on the
evolution of young neutron stars.

\end{abstract}

\section{Introduction}

The 7-s pulsar, \psr, was discovered during an automatic search of the
\asca\ archival data (Gotthelf \& Vasisht 1998; Torii \etal\ 1998).
Based on its spectral and timing properties, \psr\ is likely the latest
addition to the class of anomalous x-ray pulsars (AXPs) (Duncan \&
Thomson 1996; Mereghetti \& Stella 1995; van Paradijs \etal\
1995). Evidence included a long rotation period, large modulation
($\sim 30 \%$), steady short-term X-ray flux during the original
\ASCA\ observation, steep characteristic spectrum (power-law photon
index $\Gamma \sim 5$), location at low Galactic latitude, and
the lack of known counterpart.  A rough distance estimate derived from the
X-ray absorption places the pulsar at distance of $5-15$ kpc giving an
inferred X-ray luminosity of order $\sim 10^{35}$ erg s$^{-1}$. 

Herein we report on new \asca\ X-ray and VLA radio observations
directed at the pulsar's location. Our goal was to identify the pulsar
and confirm or repudiate the AXP hypothesis by measuring the spin-down
rate of the pulsar and searching for an associated radio supernova
remnant (SNR). We succeeded in finding a young radio SNR within the
pulsar's error circle, however the pulsator was not seen
again. Instead, we find a faint ASCA point source (Vasisht \etal\
2000) at the center of the newly discovered radio SNR \snr\ (Gaensler
\etal\ 1999).  We argue that this faint source is the pulsar \psr\ in
a low state; we consider the pulsar's location at the center of a
young SNR and the lack of a radio counterpart as evidence for the AXP
interpretation, but with a twist.

\section{THE X-RAY OBSERVATIONS}

We revisited the field containing the pulsar \psr\ with the \ASCA\
observatory on March 28-29, 1999 UT.  Figure 1 reproduced the smoothed
and exposure corrected image taken with the gas imaging spectrometers
(GISs) aboard \asca. The GIS is sensitive to photon in the $\sim 1-10$
keV energy range and has a spatial resolution $\sim 1-2^{\prime}$. All
data were edited following the standard \asca\ reduction procedures
resulting in an effective observation time is $49$~ks.

\begin{figure}[here]
\centerline{\psfig{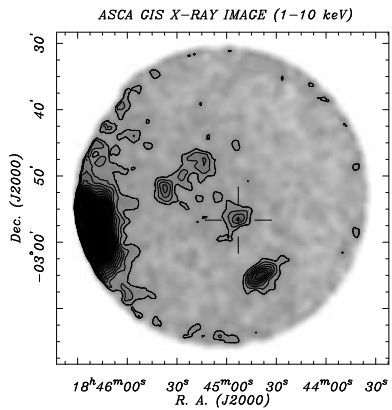}}
\caption[]{\footnotesize The follow-up \asca\ observation of \psr. The
full GIS image showing the new X-ray source \src\ marked by the cross,
and several serendipitous sources to the northeast and southwest. The
bright emission near the east edge is the SNR Kes 75.}
\end{figure}

Near the center of the field-of-view we find a faint unresolved point
source within the large $\sim 3^{\prime}$ radius error circle for
\psr.  The pulsar's poor astrometry is due to the extreme off-axis
detector location of the discovery observation. The faint source is
also detected by \asca 's solid-state imaging spectrometers (SISs) (see Fig 2)
with a similar significance of $\sim 5 \sigma$. The spatial resolution
of the SIS is $\sim 1^{\prime}$, but the derived
coordinates of $18^h 45^m 53.3^s$, $-02^{\circ} 56^{\prime}
42^{\prime\prime}$ (J2000) have an uncertainty of only $20^{\prime\prime}$
after correcting for the temperature dependent coordinate offsets
(Gotthelf \etal\ 2000).  We refer to this source as \src, and consider
whether this is indeed the expected pulsar, but at a flux level an
order of magnitude less than expected; the dearth of source photon
prohibits a proper spectral analysis or search for pulsations,
which might allow identification with \psr.

\section{THE VLA RADIO IMAGES}

Radio observations of the field of \psr\ were made with the
D-configuration of the Very Large Array (VLA) on 1999 March 26. The
total observing time was 6~hr, of which 4.5~hr was spent observing in
the 5~GHz band, and the remainder in the 8~GHz band.  At both 5 and
8~GHz a distinct shell of emission is seen, which is designated \snr\
(see Fig. 2). The shell is clumpy, with a particularly bright clump
on its eastern edge. In the east the shell is quite thick (up to 50\%
of the radius), while the north-western rim is brighter and narrower.
Two point sources can be seen within the shell interior.

The shell-like radio emission (${\sim 5\farcm0}$ in diameter) is found
to be linearly polarized and non-thermal, which, along with the lack
of significant counterpart in 60 $\mu$m {\em IRAS}\ data, are
characteristic properties of supernova remnants (e.g. Whiteoak \& Green
1996). G29.6+0.1 is thus classified as a previously unidentified SNR.
Its inferred age suggests a young remnant, with an
upper limit of 8000~yr. The location of the X-ray source
\src\ at the center of the SNR is highly unlikely to be due to a
chance superposition, suggestion that the two are related.

\begin{figure}[here]
\centerline{
\psfig{file=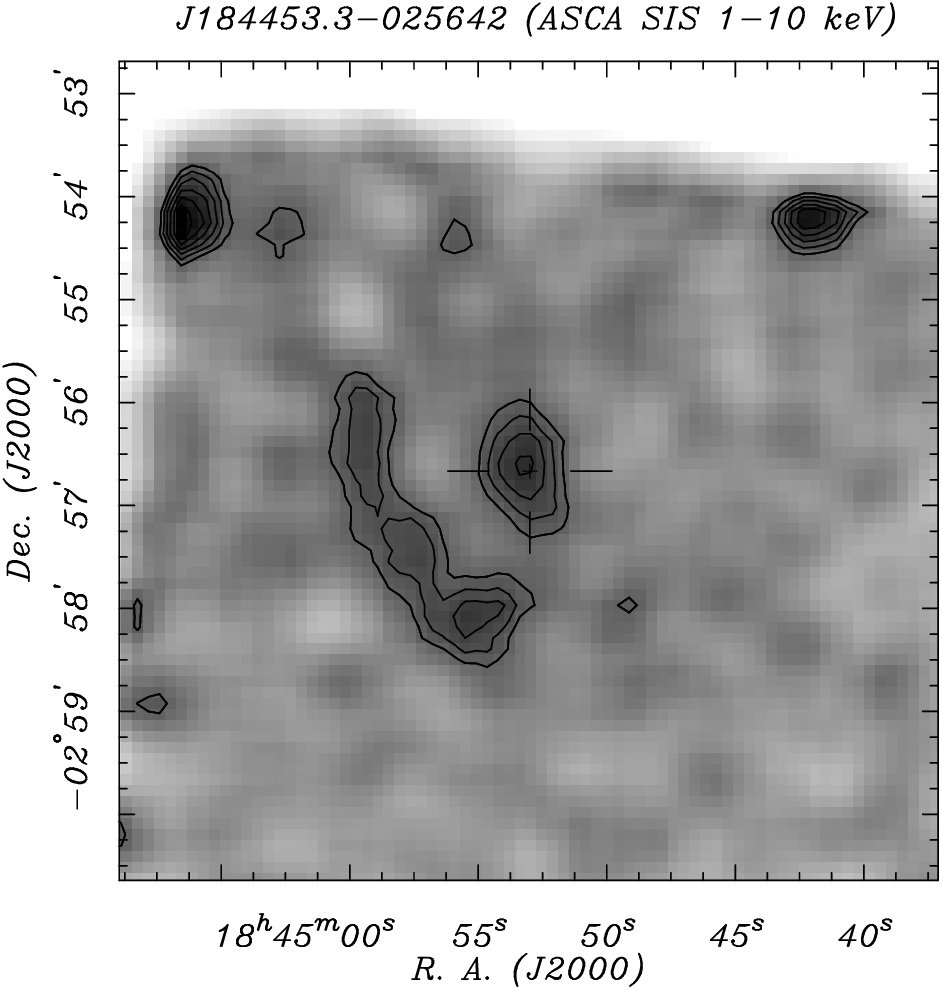,height=2.0in,angle=270,clip=} \hfil
\psfig{file=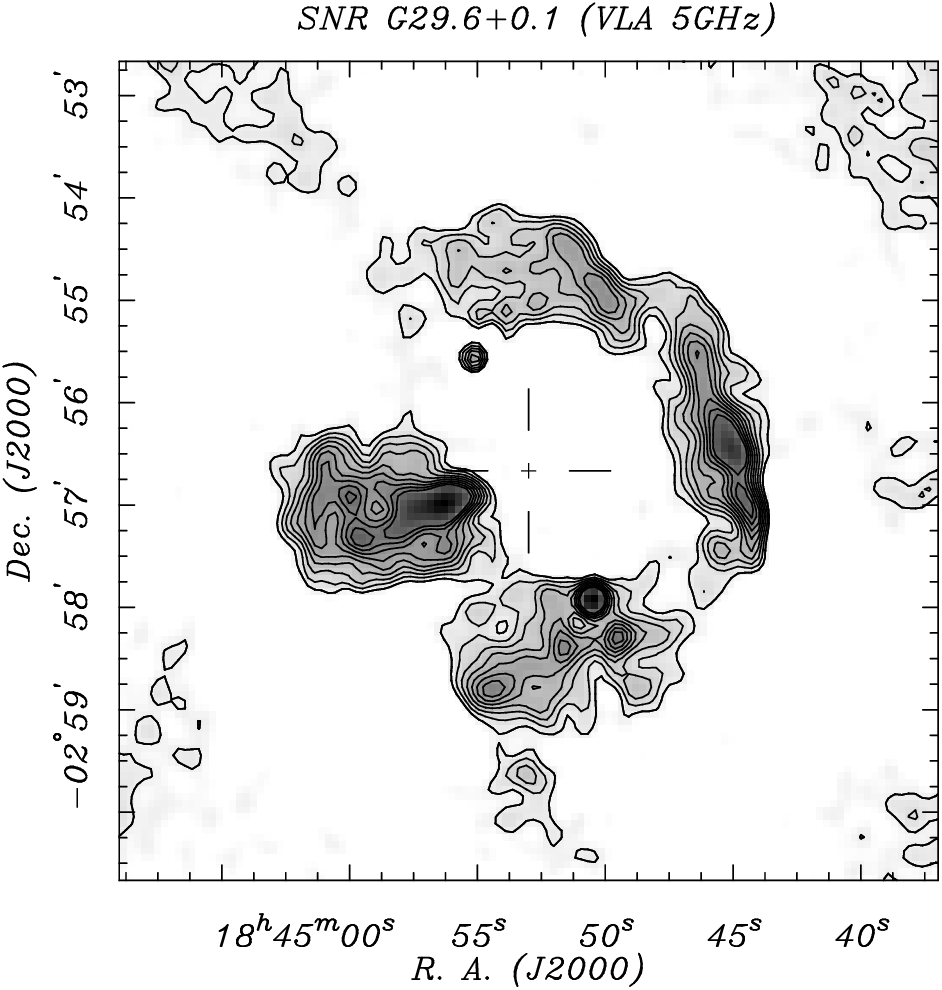,height=2.0in,angle=270,clip=}
}
\caption[]{\footnotesize Discovery of a new supernova
remnant, \snr, containing a central X-ray source, \src, within the
error box of the X-ray pulsar \psr.  ({\bf LEFT}) The \asca\ SIS X-ray
image centered on the \src, marked by the cross. An arc of emission is
evident surrounding the point source overlapping the radio shell shown in
the next panel. {\bf (RIGHT)} The 5 GHz VLA radio map of the same
region, which reveals a clumpy shell, whose spectral index is
consistent with a SNR hypothesis.   
}
\end{figure}

\section {THE NATURE OF \psr: A VARIABLE AXP?} 

The lack of a bright pulsator in the new \asca\ observation of \psr\
is quite surprising. The spectral and temporal properties of this
pulsar had strongly implied an AXP interpretation.  Indeed the
discovery of a young radio remnant coincident with the pulsar is
consistent with the AXP hypothesis. Conversely, the detection of an
X-ray point source in the center of the SNR is in itself indicative
of a neutron star candidate associated with the remnant. This new
source is exactly where we would expect the AXP to be, to within errors,
consistent with this interpretation. We therefore suggest that \src\
is indeed the pulsar, but at a much reduced ($\sim$ order of
magnitude) X-ray flux.

We now consider the interesting possibility that AXP can exhibit
extreme, factor of ten, variability. There is some evidence for this
already from two well studied AXPs which show large $\simgt 4$
variations in flux on year timescales (e.g. 1E~1048.1-593, Oosterbroek
\etal\ 1998). Most intriguing, the properties of the central,
unpulsed, neutron star candidate in SNR RCW 103 are otherwise typical
of the AXPs, but its flux has also been found to vary by an order of
magnitude at energies $>3 $ keV (Gotthelf, Petre, \& Vasisht 1999),
just what is observed for \psr. Conversely, this provides further
evidence that RCW 103 is an AXP with unseen pulsations, perhaps due to
unfavorable beaming geometry.

The identification of another AXP at the center of a young SNR have
important implications on the birth properties of pulsars. This result
is certainly consistent with AXPs being young, isolated neutron stars,
as argued by the magnetar hypothesis.  There is the possibility then
that AXPs might exhibit periods of enhanced emission. In this case,
the population of AXPs might be much greater than previously thought,
and we are only detecting a fraction of AXP, those currently in their
bright ``on'' state. Perhaps a duty-cycle (fraction of time the AXP is
``on'') of only $ \simlt 5\%$ would be required to square the known
Galactic SNRs population with the detected AXPs, if most young neutron
stars manifest themselves as AXPs as some authors suggest (see
Gotthelf 1998).  Although the magnetar hypothesis is attractive, we
cannot reject a a binary system origin, perhaps embedded and
associated with a young SNR. Further monitoring of this region is
planned.

\begin{acknowledgements}
This research is support in part by a NASA Hubble Fellowship
grant HF-01107.01-98A (B.M.G.) and by a NASA LTSA grant NAG5--7935
(E.V.G. \& G.V.).
\end{acknowledgements}

\end{document}